\documentstyle[aps,prl,multicol,epsf]{revtex}
\begin{document}
\title{\bf A Simple Model of Superconducting Vortex Avalanches}

\author{Kevin E. Bassler$^1$ and Maya Paczuski$^{1,2}$}
\address{~$^1$Department of Physics, University of Houston, Houston TX
77204-5506}
\address{~$^2$Niels Bohr Institute, Blegdamsvej 17, Copenhagen, Denmark}
\date{\today}

\maketitle 

\begin{abstract}

We introduce a simple lattice model of superconducting vortices driven
by repulsive interactions through a random pinning potential.  The
model describes the behavior at the scale of the London length
$\lambda$ or larger.  It self-organizes to a critical state,
characterized by a constant flux density gradient, where the activity
takes place in terms of avalanches spanning all length scales up to
the system size.  We determine scaling relations as well as four
universal critical exponents for avalanche moments and durations:
$\tau = 1.63 \pm 0.02$, $D=2.7 \pm 0.1$, $z=1.5 \pm 0.1$, and
$\tau_t=2.13 \pm 0.14$, for the system driven at the boundary.
\end{abstract}


\begin{multicols}{2}
\newpage
Flux penetrates hard superconductors in the form of quantized vortices
 that move via over-damped dynamics subject to repulsive interactions
 from other vortices and to random pinning forces due to
 inhomogeneities in the material.  Vortex interactions, represented by
 a modified Bessel function $K_1(r/\lambda)$, decay  with
 the London length, $\lambda$, that is much larger than either
 the size of the vortex cores or of the point pinning centers. By slowly
 increasing the external magnetic field on a thin 
 superconducting shell, vortex avalanches entering the interior have
 been observed to have a broad distribution of sizes \cite{field},
 indicating self-organized criticality \cite{soc}.
 Two dimensional molecular dynamics (MD) models \cite{jensen} of these
 experiments \cite{nori2}
indicate that vortex motion within the superconductor also
 takes place in terms of avalanches over all length scales up to the
 system size.  
 Systems of several thousand vortices corresponding to up to about
 $(30\lambda)^2$ have been numerically studied with this technique
 \cite{nori2}.

In order to explore critical behavior in the thermodynamic limit much
larger system sizes are required. For this purpose, it would be desirable to 
study a simpler model in the same universality class.  Although it is not
known to what extent universality exists in self-organized critical
(SOC) phenomena, a broad universality class encompassing one
dimensional granular piles \cite{christ}, interface depinning
\cite{leschhorn}, and earthquake models has been discovered
\cite{pbprl}.  This lends plausibility to the concept, well known in
equilibrium critical phenomena, that simple models can describe the
large scale behavior of real physical systems where the microscopic
interactions are much more complicated.  In particular, a simple model
may exist which is in the same universality class as the actual vortex system
with avalanches of all sizes.  Beyond this point, a conceptual
understanding of the interplay between repulsive vortex interactions
and random pinning would be aided by a minimal model that captures the
essential features of collective vortex dynamics.

With this view, we introduce a coarse grained lattice model to describe a vortex
 system at the scale of $\lambda$ that discards the identification of
 individual vortices, along with almost all of the microscopic degrees
 of freedom at scales smaller than $\lambda$.  
Our model is minimal, incorporating only what may be the essential
features of collective vortex dynamics:  over-damped motion of vortices,
repulsive interactions between vortices, attractive pinning interactions
at lattice defects, and to describe the experiment in
Ref. \cite{field}, boundary driving. 
One can imagine
 imposing a grid of cells on the system. In our model, vortices
 correspond to a vortex number in an extended region ($>\lambda$)
of the actual
 physical system, and the pinning corresponds to a number of
 point pins in an extended cell.  Each lattice site in our model can
 hold many vortices, and can have a different, albeit quenched,
 pinning potential, due to the underlying randomness in the positions
 and strengths of the microscopic pinning centers.  Studying our
 lattice model numerically, we can readily simulate
 much larger systems than with MD simulations, giving
 us a tool to explore scaling and phase transitions in the
 thermodynamic limit, where the system size, $L$, is large compared to
 the range of vortex interactions, $\lambda$. A
previos lattice model has been
proposed by Jensen \cite{jensenca}.  Our model differs from his 
in a number of significant ways.  The most important difference is
that 
we allow multiple vortices to occupy each site, consistent with
the coarse graining idea \cite{jencomment}.  

We find that our simple model exhibits self-organized criticality.
The observed critical exponents are universal in the sense that they
do not vary over a range of parameter values in the model.  We drive
our system in a manner that represents the experiments of Field {\it
et al} \cite{field}.  As vortices are slowly pumped in at the left end
of our system and allowed to leave at the right end, the gradient of
the vortex density builds up to a constant value throughout the
system, in agreement with the picture of Bean \cite{bean}.  The vortex
model acts like a pile of sand!  It is important to note that the
vortex pile is not minimally stable (which is a local criterion first
suggested by Bean), but it is marginally stable \cite{soc}.  The 
minimally stable
state is unstable to plastic deformations or avalanches.
  The actual condition of criticality is
a global one where no length scale other than the system size plays
any role whatsoever.  We find a temporal pattern of intermittent
bursts of vortices leaving the system, as well as internal avalanches.
We apply finite size scaling methods to the histograms for the sizes $s$
and duration $t$ of avalanches, determining four critical
exponents, which agree with known scaling relations for a boundary
driven system.
The values of the
critical exponents found are close to those of the two dimensional
``linear'' interface model \cite{leschhorn},  suggesting a common
universality class.

\begin{figure}
\narrowtext
\epsfxsize=3.0truein
\hskip 0.05truein \epsffile{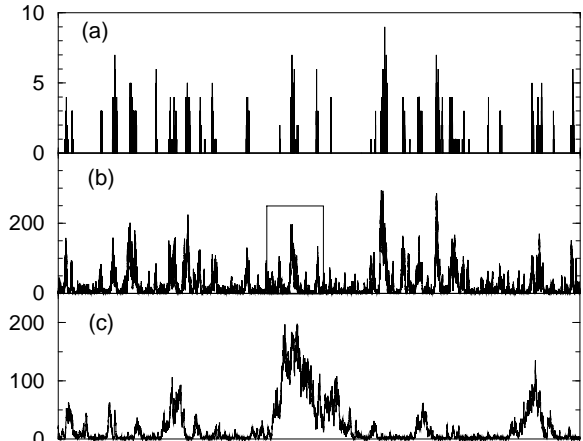}
\caption{ Time series of the vortex dynamics.  Frame (a) shows the
number of vortices falling off the system. Frame (b) shows the number
of moving vortices, i.e.\ the activity.  Frame (c) shows a
magnification of the boxed region of (b). Both (a) and (b) show $10^5$
time steps.  }
\end{figure}

Our model is a coarse-grained representation of the microscopic vortex
 dynamics in which the force on a vortex $i$ is given by given by the
 over-damped equation of motion $\mbox{{\bf f}}_i = \mbox{{\bf
 f}}_{int}+\mbox{{\bf f}}_{pin} \propto \mbox{{\bf v}}_i$, where
 $\mbox{{\bf f}}_{int}$ is the sum of the repulsive forces from other
 vortices, $\mbox{{\bf f}}_{pin}$ is the sum of the attractive forces
 from the pinning centers, and $\mbox{{\bf v}}_i$ is
the velocity of vortex $i$ \cite{jensen,nori2}.  We consider a
 two dimensional honeycomb lattice.  Each lattice site $x$ is occupied
 by an integer $m(x)$ vortices, and has three nearest neighbors.
 Vortices repel others occupying the same lattice site and, more
 weakly, those on nearest neighbor sites. There is also an attractive
 pinning force at random sites.  As in the microscopic case,
 both of these coarse-grained forces are the gradient of a potential.
 The force on a vortex at a site $x$ in the direction of a nearest
 neighbor site $y$ is calculated by taking a discrete gradient of the
 sum of those two potentials,
\begin{eqnarray}
F_{x \rightarrow y} =  - V_{pin}(x) + V_{pin}(y) +
\left[ m(x) - m(y) - 1 \right] \nonumber \\ 
+  r\left[ m(x1) + m(x2) - m(y1) - m(y2) \right]   \quad  ,
\end{eqnarray}
where $y$, $x1$, and $x2$ are the nearest neighbors of $x$, and
$x$, $y1$, and $y2$ are the nearest neighbors of $y$.
The scale of the problem is set so that strength of the on-site
force is unity, and strength of the nearest neighbor force is $r<1$.
The strength of the quenched random pinning potential at $x$, $V_{pin}(x)$, is chosen to be $p$
with probability $q$, and 0 with probability $1-q$.
Thus, the density of pinning centers is $q$, and
there are three parameters in the model: $r$, $p$, and $q$.
In the figures shown here we use the parameters $(r,p,q) = (0.1,5.0,0.1)$.  
We have also simulated systems
with $(r,p,q) = (0.2,5.0,0.1), \; (0.1,5.0,0.4)$, and $(0.1,1.0,0.1)$.
In all four cases, the critical exponents remained unchanged
within numerical errors.

A vortex moves one lattice site in a direction
 when the force in that direction is greater than zero. 
Thus, even sites with $V_{pin}=0$ can pin vortices.
If there is more than one unstable direction for a vortex to move in, 
one direction is picked at random.
All lattice sites are updated in parallel, and
only one vortex can move from each site on a particular update.
As in experiments, any vortex that reaches the right edge
of the system is removed. 
Vortices are forbidden to move off the left edge of the system \cite{explain}.
Periodic boundary conditions apply at the top and bottom of the lattice.
We use approximately square systems so that the avalanches do not
wrap around on themselves.

\begin{figure}
\narrowtext
\epsfxsize=2.5truein
\hskip 0.25truein
\epsffile{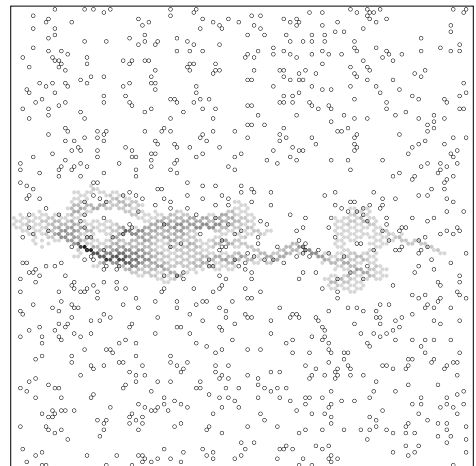}
\caption{ Greyscale plot showing an avalanche. The filled
circles indicate sites from which a vortex has moved during the
avalanche, where the darkness indicates the number of topplings.
 The open circles represent strong pinning
centers.}
\end{figure}

An avalanche is initiated by adding a vortex to a stable configuration at a
randomly chosen site on the left, or loading, edge of the lattice.
The avalanche continues by repeatedly updating the lattice until there
are no longer any unstable directions for a vortex at any lattice
site, and the configuration is again stable.  The limit of slow
driving is achieved by adding a vortex to the loading edge only after
the previous avalanche has ended.

Initially, the vortex pile is empty.  As it fills up with vortices,
eventually a chain reaction of sliding events leads to one or more
vortices leaving the system at the output edge.  The global cascade of
events (topplings) is due to the local repulsive interactions of
vortices.  An avalanche constitutes a type of generalized branching
process because a toppling at one site can only affect its nearest and
next nearest neighbors at the next time step.  This fact allows an
efficient list algorithm to be used to simulate the system,
checking only sites on the list of possibly active sites for instability.
Eventually for every vortex added, one on average leaves the system.
In this stationary state, the average vortex density acquires a
constant gradient throughout the system.  The vortices form a vortex
pile, with fluctuations about an average slope much like a sand pile.

In Fig. 1, we show a time series of the activity in the stationary state
for a  system with
$L=200$.  The activity takes place intermittently with bursts over all
temporal scales up to a cutoff that grows with system size.
  When a given bursts ends, at a time $t$,
the activity $n(t)$ reaches zero
and another vortex is added at the loading edge.  Fig. 1(a) shows
the number of vortices that fall off the right edge as
a function of time.  It has a similar qualitative appearance to
that seen in experiments \cite{field}.  Figs. 1(b) and 1(c)
show the number of moving vortices as a function of time.
These signals are qualitatively similar to what is seen in MD simulations \cite{nori2}.
Within each avalanche there are ``bursts within bursts''.

The spatial behavior of an avalanche in the stationary state
is shown in Fig. 2.  The number of topplings at a given site
is represented by a grey scale.  Qualitatively, the avalanche has
an inhomogeneous behavior, with holes where there is no activity inside
a region of activity.  One can characterize an avalanche
by the maximum extent of its penetration in the direction of flow, $r$,
 the total number of topplings
in the avalanche, $s$, its duration, $t$, and $n(t)$, among other quantities. 
Because the mass of the avalanche
represents the time integral over the sequence of topplings, 
 $s\sim n t$, as in other models of SOC. The scaling dimensions of these
variables are defined for avalanches as $s\sim r^D$ and $t \sim r^z$
\cite{pmb}.

In analogy with other models of SOC \cite{pmb}, as well as other
critical phenomena,
we use the following scaling ansatz for the probability distribution
$P(s,L)$ to have an avalanche of size $s$ in a system of linear extent
$L$:
\begin{equation}
P(s,L)= s^{-\tau}g(s/L^D) \qquad .
\end{equation}
This scaling ansatz is confirmed for our vortex model by numerical
simulations as shown in Fig. 3.  The avalanche dimension, $D$, and the
distribution exponent, $\tau$, are not independent for our boundary
driven system.  In the stationary state, every row, on average, must
transfer one vortex to the right.  In this case, the average avalanche
size size $<s>=cL$ where $c$ is some constant. We find $c=1.1$, 
slightly larger than one since some
toppling events occur counter to the average flow.  (For the one
dimensional Oslo model \cite{christ} $c=1$ exactly \cite{pbprl}.) We
have checked to see that the actual moment of the sliding has the same
dimension as the total number of topplings; so that the back topplings
are not significant.  The requirement $<s>\sim L$ leads to $\tau= 2 -
1/D$.  We align the cutoff regime in the data collapse plot to
determine $D=2.7 \pm 0.1$ and choose
$\tau=1.63 \pm 0.02$ to give a flat plateau in the data
collapse  over four orders of magnitude in $s/L^D$.  These values
agree perfectly with the above scaling relation.

\begin{figure}
\narrowtext
\epsfxsize=3.0truein
\hskip 0.05truein \epsffile{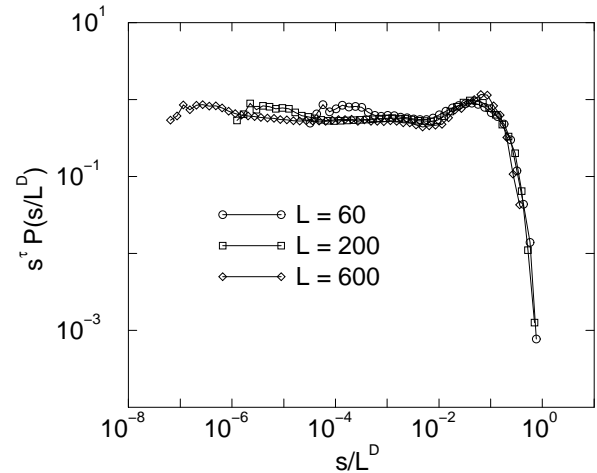}
\caption{
Finite size scaling plot of the avalanche size histograms.
 Each curve is calculated
from $10^7$ avalanches on each of 
five (two for the $L=600$ system) different 
realizations of the quenched disorder. The exponents used
 are $D = 2.70$ and $\tau = 1.63$.
}
\end{figure}

A similar scaling ansatz is used for the probability distribution
$P(t,L)$ to have an avalanche of duration $t$ in a system of linear
extent $L$:
\begin{equation}
P(t,L)= t^{-\tau_t}g(t/L^z) \qquad ,
\end{equation}
where the dynamical exponent $z$ determines the system size cutoff
in durations $t_{co} \sim L^z$.  Conservation of probability gives
an additional scaling relation $D(\tau -1)=z(\tau_t -1)=\tau_r -1$,
where $\tau_r$ is the histogram exponent for the
spatial penetration of avalanches \cite{pmb}.  Here $\tau_r=D$.
Since for every avalanche, its
size is greater than or equal to its duration, $\tau_t > \tau $ 
and $D\geq z$.
In Fig. 4, we use the position of the cutoff region to find $z=1.5 \pm 0.1$
 and choose $\tau_t = 2.13 \pm 0.14$ to give a flat plateau in the
data collapse.  These values obey the scaling relation above.

Note that the values for $D$ and $z$ are very close to those measured
for the ``linear'' interface depinning model in two dimensions, where
$D=2.75 \pm 0.05$ and $z=1.58 \pm 0.04$ \cite{leschhorn,pmb}. The
histogram exponents differ when these models are driven uniformly,
rather than at the boundary. For uniform driving
$D(2-\tau)=2$.  

The distribution of fall off events can be determined by measuring the
 total number of vortices, $f$, that leave the system in each avalanche.
 This distribution is broad up to a cutoff $f_o(L)$ that diverges
 with system size as $f_o \sim L^{D-1}$, but it does not appear to
 show power law behavior for $f < f_o$.  However, since the system is
 driven at the opposing boundary, only the few avalanches that 
cross the system make any vortices fall off the output edge.
 Thus the statistics for the
 fall off data are not as good as that of the quantities
 characterizing internal avalanches.  Also, to our
 knowledge, there is no reason to expect that the fall
 off distribution should be a power law, even though the internal
 avalanche size distribution is itself a power law.  Experiments have
 measured fall off distributions from time series involving about
 $10^4$ avalanches in a system with $L/\lambda \approx 10^3$.  Both
 the value of the cutoff and the apparent power law appeared to depend on the
 magnitude of the external magnetic field.   This dependence
could be consistent with our results.  The problem of
 interpreting fall off data exists in many other SOC systems as
 well \cite{rice}.

\begin{figure}
\narrowtext
\epsfxsize=3.0truein
\hskip 0.05truein
\epsffile{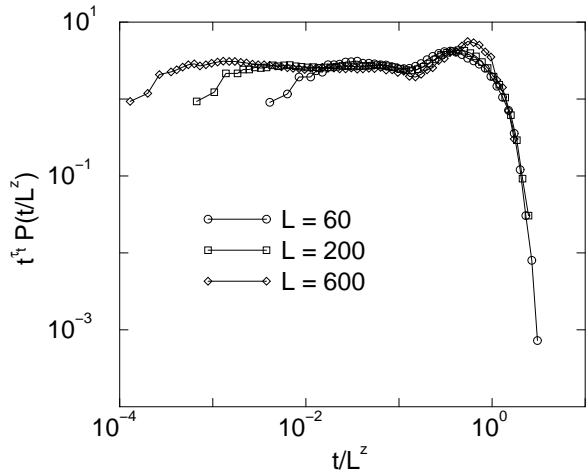}
\caption{ Finite size scaling plot of the avalanche duration
histograms.  The exponents used are $z = 1.50$ and
$\tau_t = 2.13$.}
\end{figure}

Ideally, experiments to measure internal avalanches could be devised. For example,
Yeh and Kao \cite{yeh} attached a SQUID to a superconducting sample and were
able to measure fluctuations associated with the flux flow. In order
to compare with our results an array of SQUIDs could be used. 
Alternatively, it may eventually be possible using Lorentz
microscopy techniques \cite{matsuda}. MD simulations do provide
some points of comparison for scaling behavior of internal
avalanches.  Those results, however, have not been cast using
collapse techniques due, presumably, to the finite size limitations in
$L/\lambda$.  In any event, the exponent $\tau$ in Ref. \cite{nori2}
varied for different parameter
values over a range $1.1 \leq \tau \leq 1.7$.  We suspect that this
variation is  a finite size effect.  Nevertheless our value
$\tau = 1.63$ falls within this range \cite{time}.  We propose that the scaling
relation for the average moment $<s>\sim L$ would also hold for the MD
simulations, and for the real physical system, as well.

Although the SOC behavior we observe in the vortex lattice model is
robust, having universal critical exponents over a range of
parameters, it is possible that the behavior changes for parameters outside 
the range we have reported here. For example, for a sufficiently rugged pinning
landscape, the vortex motion may be locked into isolated channels,  
introducing a scale in the avalanche distribution \cite{nori2,pla},
consistent with the experiments of Zieve {\it et al} \cite{zieve}. 
Similarly, for a 
sufficiently 
flat pinning potential the avalanches become very wide and may have
a characteristic size as well. We are currently investigating both of these
situations. 
It would be interesting to study other properties of vortex dynamics using our
model.  These include magnetic relaxation
\cite{aegerter}, hysteresis, aging, using a vortex weakened pinning
potential to take into account interstitial pinning, driving the
system at a finite rate, or in a different manner that would
correspond to applying an electrical current.

We thank G. Reiter, F. Nori, and P. Bak for helpful discussions.

\end{multicols}

\begin{references}
\bibitem{jencomment}Also, our model driven in a different way,
with a different dynamical update rule.
\bibitem{explain}
 The forces on the vortices at sites on the left edge of the lattice are 
calculated by
assuming that their nearest neighbor sites just off the lattice are
occupied by the same number of vortices as that site had the last time a new
vortex was added to the lattice at that site.
\bibitem{time}
It is important to note, however, that the distribution
of avalanche lifetimes do not compare well. Ref.\ [4] reports a critical
exponent for $\tau_t=1.4$ ($\gamma=1.4$ in their notation) which
not only is smaller than some of their reported values for $\tau$,
violating $\tau_t>\tau$, but also considerably smaller than our
$\tau_t=2.13$.  Note that our exponent $\tau$ corresponds to $\delta$
in Ref.\ [4].
\end{references}
\end{document}